\documentclass[prl,twocolumn,showpacs]{revtex4-1}
\usepackage{bm,amsmath,amssymb}
\usepackage{graphicx}
\usepackage{subfigure}
\usepackage{color}
\usepackage{multirow}

\newcommand \ber {\begin{eqnarray}}
\newcommand \eer {\end{eqnarray}}
\newcommand \beq {\begin{equation}}
\newcommand \eeq {\end{equation}}

\begin{document}

\title{Charm Degrees of Freedom in the Quark Gluon Plasma}

\author{Swagato Mukherjee, Peter Petreczky and Sayantan Sharma}

\affiliation{Physics Department, Brookhaven National Laboratory, Upton, NY 11973,
USA}

\begin{abstract}

Lattice QCD studies on fluctuations and correlations of charm quantum number have
established that deconfinement of charm degrees of freedom sets in around the chiral
crossover temperature, $T_c$, {\sl i.e.} charm degrees of freedom carrying fractional
baryonic charge start to appear. By reexamining those same lattice QCD data we
show that, in addition to the contributions from quark-like excitations, the partial
pressure of charm degrees of freedom may still contain significant contributions from
open-charm meson and baryon-like excitations associated with integral baryonic
charges for temperatures up to $1.2~ T_c$. Charm quark-quasiparticles
become the dominant degrees of freedom for temperatures $T>1.2~ T_c$.

\end{abstract}

\maketitle

Nuclear modification factor and elliptic flow of open-charm hadrons in heavy-ion
collision experiments are important observables that provide us with detailed knowledge of
the strongly coupled quark gluon plasma (QGP) \cite{Rapp:2009my}. Most of the
theoretical models that try to describe these quantities rely on the energy loss of
heavy quarks via Langevin dynamics \cite{Moore:2004tg,Berrehrah:2014kba,Cao:2015hia}.
However, the importance of possible heavy-light (strange) bound states inside QGP
have been pointed out in Refs.
\cite{He:2012df,He:2014cla,Adil:2006ra,Sharma:2009hn}. In particular, presence of
such heavy-light bound states above the QCD transition temperature seems to be
necessary for the simultaneous description of elliptic flow and nuclear modification
factor of $D_s$ mesons \cite{He:2012df}. Presence of various hadronic bound states
\cite{Ratti:2011au} as well as colored \cite{Shuryak:2003ty,Shuryak:2003xe} ones in
strongly coupled QGP created in heavy-ion collisions have also been speculated in
other other contexts. 

By utilizing various novel combinations of up to fourth order cumulants of
fluctuations of charm quantum number ($C$) and its correlations with baryon number
($B$), electric charge and strangeness ($S$) lattice QCD studies
\cite{Bazavov:2014yba} have established that charm degrees of freedom 
associated with fractional baryonic and electric charge start appearing at the chiral crossover temperature,
$T_c=154\pm9$ MeV \cite{Bazavov:2011nk,Borsanyi:2010bp,Bhattacharya:2014ara}. Below $T_c$
the charm degrees of freedom are well described by an uncorrelated
gas of charm hadrons having vacuum masses \cite{Bazavov:2014yba}, {\sl i.e.} by
the hadron resonance gas (HRG) model. Similar conclusions were also obtained from
lattice QCD studies involving the light up, down and strange quarks
\cite{Bazavov:2013dta, Bazavov:2014xya}. 

\begin{figure}[t!]
\begin{center}
\includegraphics[width=0.48\textwidth,height=0.25\textheight]{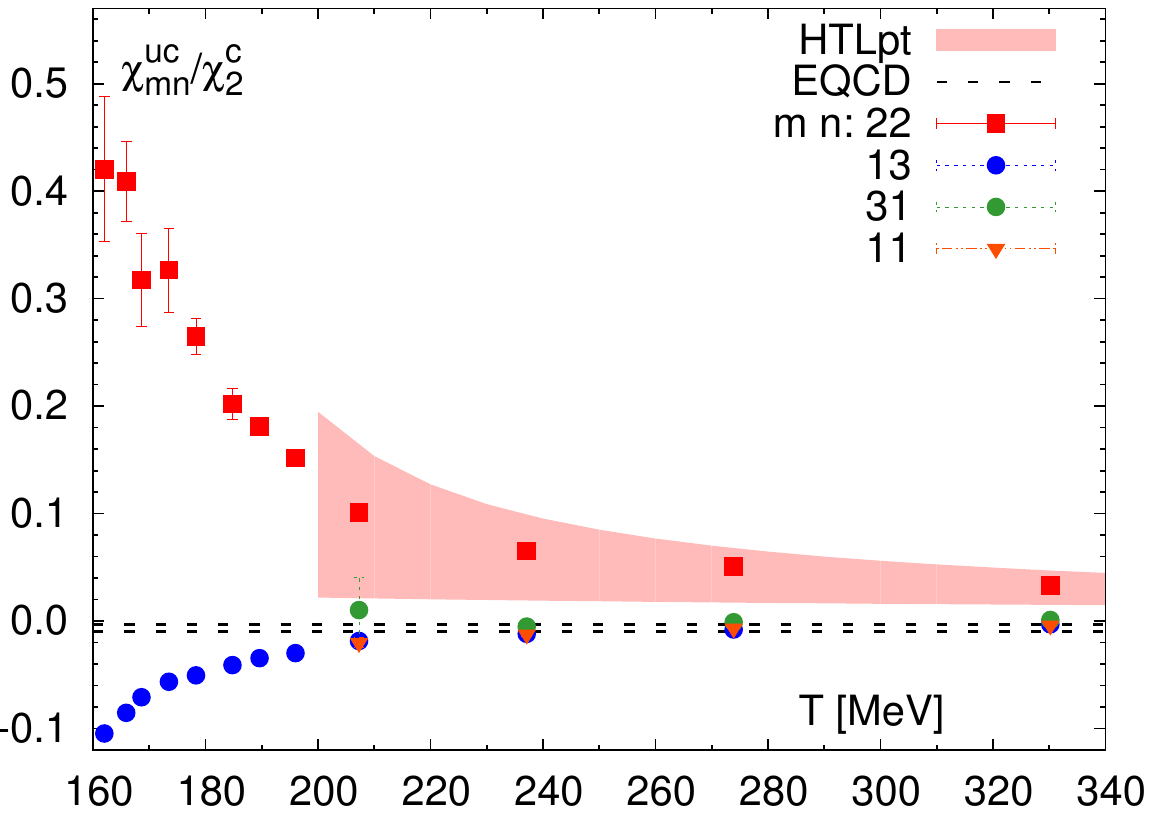}
\end{center}
\caption{Off-diagonal quark number susceptibilities $\chi_{nm}^{uc}$ normalized by
the second order diagonal charm susceptibility $\chi_2^c$ as a function of
temperature \cite{Bazavov:2014yba}. The shaded band shows the three loop hard thermal
loop perturbation theory calculation for $\chi_{22}/\chi_2$; the width of the band
corresponds to a variation of the renormalization scale from $\pi T$ to $4 \pi T$
\cite{Haque:2014rua}. Also shown, as dashed lines, are the results of dimensionally
reduced EQCD calculations for $\chi_{11}$ corresponding to temperatures $1.32~T_c$ and
$2.30~T_c$ from \cite{Hietanen:2008xb}.}
\label{fig:nondiag}
\end{figure}

On the other hand, lattice QCD calculations have also shown that weakly interacting
quasi-quarks are good descriptions for the light quark degrees only for temperatures
$T\gtrsim2~T_c$ \cite{Bazavov:2013dta,Bazavov:2013uja,Ding:2015fca,Bellwied:2015lba}.
The situation for the heavier charm quarks is also analogous. By re-expressing the
lattice QCD results for charm fluctuations and correlations up to fourth order from 
Ref. \cite{Bazavov:2014yba} in the charm ($c$) and up ($u$) quark flavor basis, we show 
the $u$-$c$ flavor correlations, defined as  $\chi_{mn}^{uc}=(\partial^{m+n}p/
\partial\hat\mu_u^m\partial\hat\mu_c^n)$ at $\mu_u=\mu_c=0$ in
Fig. \ref{fig:nondiag}. Here, $p$ denotes the total pressure in QCD, $\mu_u$ and $\mu_c$ indicate the
up and charm quark chemical potentials with $\hat\mu_X\equiv\mu_X/T$. In order to
compare these lattice QCD data with resummed perturbation
theory results, which are available only for zero quark masses, we normalize the
off-diagonal flavor susceptibilities with the second order charm quark
susceptibility $\chi_2^c=(\partial^2p/\partial\hat\mu_c^2)$ calculated at $\mu_X=\mu_c=0$. Such
a normalization largely cancels the explicit charm quark mass dependence of the
off-diagonal susceptibilities and enables us to probe whether the $u$-$c$ flavor
correlations can be described by the weak coupling calculations. In the weak coupling
limit $\chi_{11}^{uc},~\chi_{13}^{uc}$ and $\chi_{31}^{uc}$ are expected to have leading 
order contributions at ${\cal O}(\alpha_s^3)$ \cite{Blaizot:2001vr}, where $\alpha_s$ is
the QCD strong coupling constant. This contribution is, strictly speaking,
non-perturbative but can be calculated on the lattice using 
dimensionally reduced effective theory for high temperature QCD, the so-called
electrostatic QCD (EQCD) \cite{Hietanen:2008xb}.  Similarly, in the weak coupling picture,
the leading contribution to $\chi_{22}^{uc}$ arises from the so-called plasmon term
and starts at ${\cal O}(\alpha_s^{3/2})$ \cite{Haque:2014rua}. Thus, it is generically expected
$\chi_{22}^{uc}\gg\chi_{13}^{uc}\sim\chi_{31}^{uc}\sim\chi_{11}^{uc}$ in the weak coupling limit. As shown in
Fig. \ref{fig:nondiag} such an obvious hierarchy in magnitude of the off-diagonal
susceptibilities is clearly absent in the lattice data for $T<200$ MeV. However, for
$T\gtrsim200$ MeV these lattice results are largely consistent with the weak coupling
calculations, indicating that the weakly coupled quasi-quarks can be considered as
the dominant charm degrees freedom only above this temperature. The fact that for
$T_c\lesssim T\lesssim 200$ MeV the charm degrees of freedom are far from weakly
interacting quasi-quarks is also supported by lattice QCD studies of the screening
properties of the open-charm mesons. In this temperature range  the screening masses
of open-charm mesons also turn out to be quite different from the expectation based on
an uncorrelated charm and a light quark degrees of freedom \cite{Bazavov:2014cta}.

From the preceding discussion it is clear that the weakly interacting charm
quasi-quarks cannot be the only carriers of charm quantum number for $T\leq 200$ MeV.
Such an observation naturally raises the question whether charm excitations associated with baryon number zero
and one, exist in QGP for $T_c\lesssim T\lesssim 200$, along with the charm
quasi-quark excitations carrying $1/3$ baryonic charge. In the present work, we
address this question by postulating that such open-charm meson and baryon-like
excitations exist alongside the charm quasi-quarks in QGP and then investigate
whether such an assumption is compatible with the exact lattice QCD results on charm
fluctuations and their correlations.

Charm fluctuations and their  correlations with other conserved quantum numbers can be measured on
the lattice through the generalized charm susceptibilities
\begin{equation}
\chi_{ijk}^{XYC} = \left. \frac{\partial^{i+j+k} p(T,\mu_X,\mu_Y,\mu_C)}
{\partial\hat\mu_X^i\partial\hat\mu_Y^j\partial\hat\mu_C^k}
\right\vert_{\mu_X=\mu_Y=\mu_C=0} ,
\end{equation}
where $\hat\mu_X=\mu_X/T$. For notational brevity we will suppress the superscripts
of $\chi$ whenever the corresponding subscript is zero. To check our postulates
against the lattice QCD results, throughout this study, we will use the lattice QCD
data of Ref.  \cite{Bazavov:2014yba} on up to fourth order generalized charm
susceptibilities, {\sl i.e.} for $i+j+k\leq4$.

To avoid introduction of unknown tunable parameters we simply postulate an
uncorrelated, {\sl i.e.} non-interacting gas of charm meson, baryon and quark-like
excitations for $T\gtrsim T_c$. Owing to the large mass of the charm quark itself, compared to
$T\sim2~T_c$, it is safe to treat all the quark, meson and baryon-like
excitations as classical quasi particles, {\sl i.e.} within the Boltzmann
approximations. Furthermore, as discussed in Ref. \cite{Bazavov:2014yba}, the doubly
and triply charmed baryons are too heavy to have any significant contributions to QCD
thermodynamics in the temperature range of interest and we thus neglect their
contributions. With these simplifications the partial pressure of the open-charm
sector, $p^C$, can be written as
\begin{eqnarray}
& p^C (T,\mu_C,\mu_B) = p_q^C(T) \cosh\left(\hat\mu_C+\hat\mu_B/3\right) + 
\nonumber \\
& ~~~~~ p_B^C(T) \cosh\left(\hat\mu_C+\hat\mu_B\right)
+ p_M^C(T) \cosh\left(\hat\mu_C\right) \,,
\label{eqn:pc}
\end{eqnarray}
where $p_q^C$, $p_B^C$ and $p_M^C$ denote the partial pressure of the quark-like,
meson-like and baryon-like excitations, respectively, and $\mu_B$ and $\mu_C=\mu_c$
represents the baryon and charm chemical potentials. 

\begin{figure}[!t]
\begin{center}
\includegraphics[width=0.48\textwidth,height=0.45\textheight]{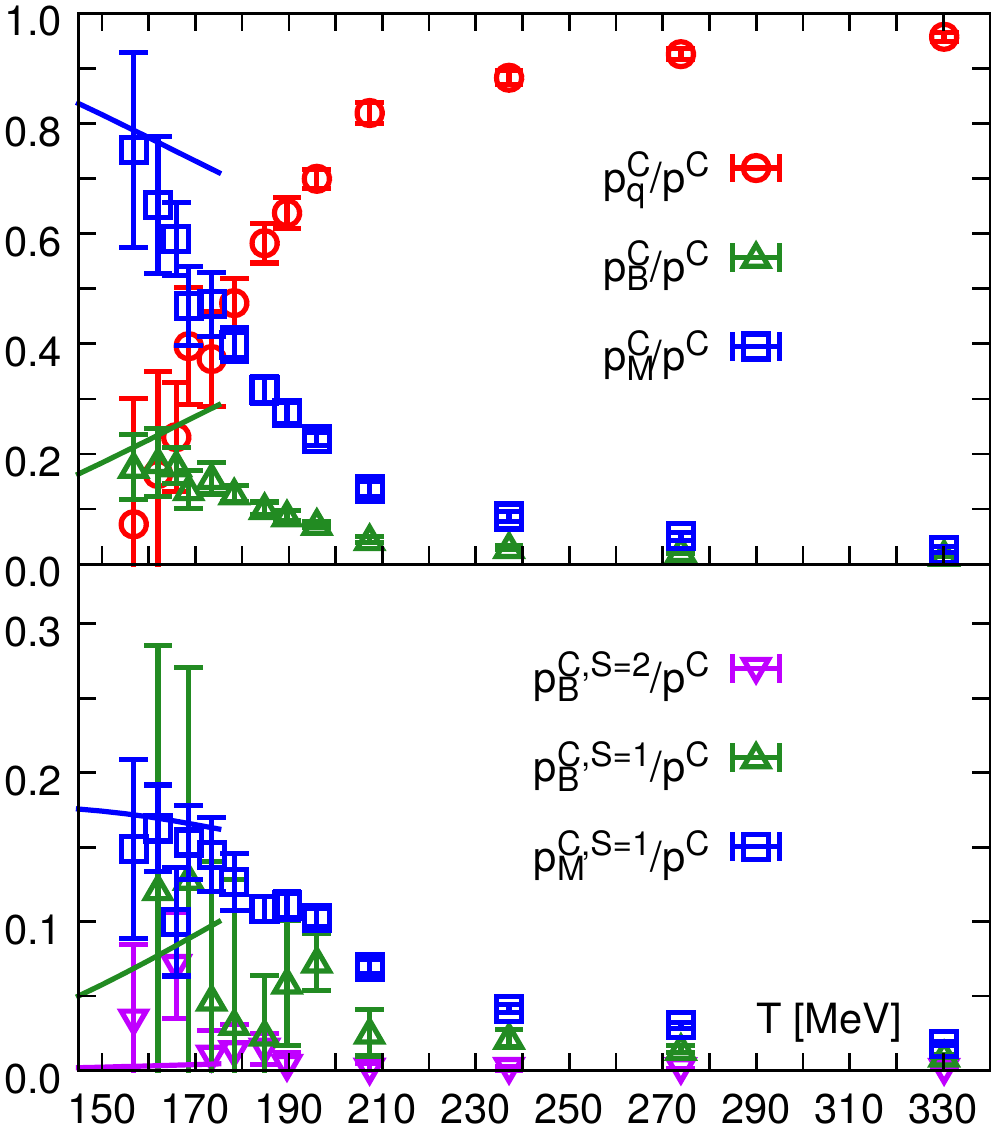}
\end{center}
\caption{(Top) Fractional contributions of partial pressures of charm quark-like
($p_q^C$), meson-like ($p_M^C$), and baryon-like ($p_B^C$) excitations to the total
charm partial pressure ($p^C$). (Bottom) Fractional contributions of partial
pressures of charm-strange meson-like ($p_M^{C,S=1}$), charm-singly-strange
baryon-like ($p_B^{C,S=1}$) and charm-doubly-strange baryon-like ($p_B^{C,S=2}$)
excitations to the total charm partial pressure ($p^C$). The solid lines show the
corresponding partial pressures obtained from HRG model including additional quark
model predicted charm hadrons (see text).}
\label{fig:pci}
\end{figure}

Using combinations of up to fourth order baryon-charm susceptibilities it is easy to
isolate the partial pressures of $p_q^C$, $p_M^C$ and $p_B^C$ appearing in Eq.
\ref{eqn:pc}. For example, $p_q^C=9(\chi_{13}^{BC}-\chi_{22}^{BC})/2$,
$p_B^C=(3 \chi_{22}^{BC}-\chi_{13}^{BC})/2$, and
$p_M^C=\chi_2^C+3\chi_{22}^{BC}-4\chi_{13}^{BC}$. The contributions of these
partial pressures compared to total charm pressure $p^C(T,0,0)=\chi_2^C$ is shown in Fig.
\ref{fig:pci} (top). For $T\lesssim T_c$ the partial pressure of mesons, $p_M^C$ and 
the partial pressure of baryons, $p_B^C$ agree with the
corresponding partial pressures from the HRG model including all the experimentally
observed as well as additional quark model predicted but yet unobserved open-charm
hadrons with vacuum masses \cite{Bazavov:2014yba}. The contributions of $p_M^C$ and
$p_B^C$ remain significant till $T\lesssim 200$ MeV. In fact, for $T\lesssim180$ MeV
the combined contributions of $p_M^C$ and $p_B^C$ exceeds the contribution from
$p_q^C$. With increasing temperatures $p_M^C$ and $p_B^C$ deviate from the HRG model
predictions. This indicate that these charm meson and baryon-like excitations can no
longer be considered as vacuum charm mesons and baryons. This is in line with the
lattice QCD studies on spatial correlation functions of open-charm mesons
\cite{Bazavov:2014cta}, which show significant in-medium modifications of open-charm
mesons already in the vicinity of $T_c$. The partial pressure of quark-like
excitations is quite small for $T\sim T_c$ and becomes the dominant contribution to
$p^C$ only for $T>200$ MeV. 

Since a charm-quark-like excitation does not carry a strangeness quantum number, the
excitations carrying both strangeness and charm quantum numbers are a much cleaner probe of the
postulated existence of the charm hadron-like excitations. In this sub-sector, the
pressure can be partitioned into partial pressures of $|C|=1$ meson-like excitations
carrying strangeness $|S|=1$ and $C=1$ baryon-like excitations with $|S|=1,2$, {\sl
i.e.}
\begin{eqnarray}
& p^{C,S}(T,\mu_B,\mu_S,\mu_C) = 
p_{M}^{C,S=1}(T) \cosh\left(\hat\mu_S+\hat\mu_C\right) + \nonumber \\
& \sum_{j=1}^{2} p_B^{C,S=j}(T) \cosh\left(\mu_B-j\mu_S+\mu_C\right) .
\end{eqnarray}
Thus, the partial pressures of the strange-charm hadron-like excitations can be
obtained as: $p_M^{C,S=1}=\chi_{13}^{SC}-\chi_{112}^{BSC}$,
$p_B^{C,S=1}=\chi_{13}^{SC}-\chi_{22}^{SC}-3 \chi_{112}^{BSC}$, and
$p_B^{C,S=2}=(2\chi_{112}^{BSC}+\chi_{22}^{SC}-\chi_{13}^{SC})/2$. In Fig.
\ref{fig:pci} (bottom) we show the fractional contributions of these partial
pressures towards the total charm partial pressure $p^C(T)=\chi_2^C$.  Even in this
sub-sector, contributions from the hadron-like excitations are significant for $T\lesssim200$
MeV. However, partial pressure for the $S=2$ charm baryon-like excitations is
negligible.     

\begin{figure}[!t]
\begin{center}
\includegraphics[width=0.48\textwidth,height=0.25\textheight]{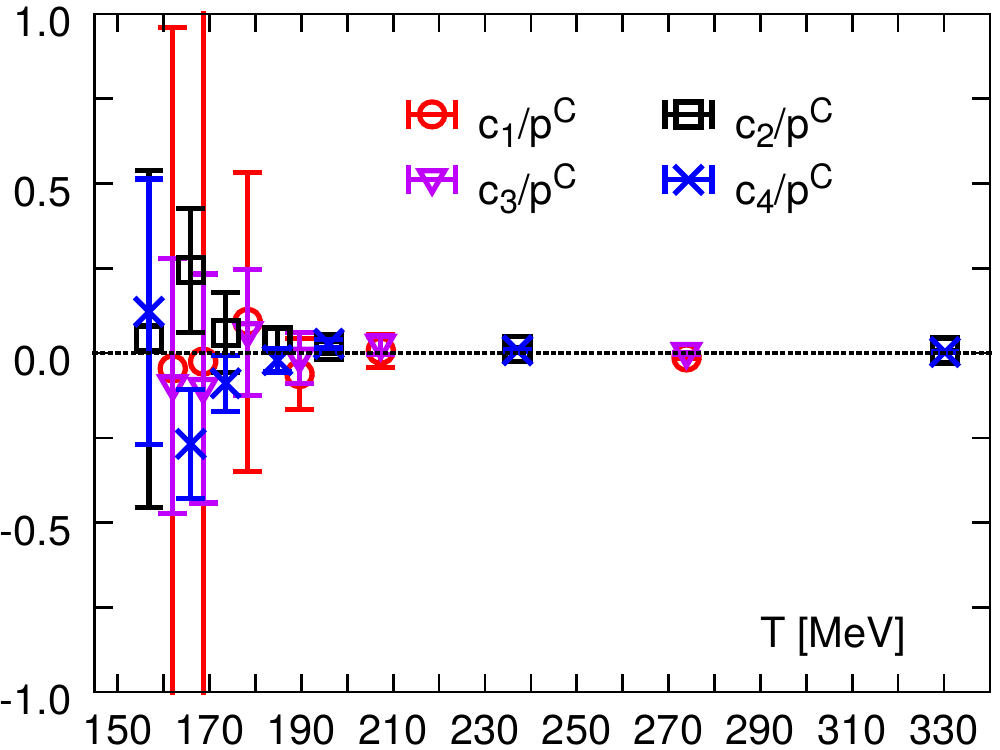}
\end{center}
\caption{Lattice QCD results for four constraints ($c_i$) normalized by the total charm pressure (see text).}
\label{fig:const}
\end{figure}

Having shown that there can be significant contributions from charm meson and
baryon-like excitations to the charm partial pressure in QGP, it is important to ask
whether the addition of only these charm degrees of freedom besides the charm
quark-like excitations is sufficient to describe all available lattice QCD
results for up to fourth order charm susceptibilities. As discussed previously in
Ref. \cite{Bazavov:2014yba},  the constraints $\chi_4^C=\chi_2^C$, $\chi_{11}^{BC}=\chi_{13}^{BC}$,
$\chi_{11}^{SC}=\chi_{13}^{SC}$ are due to negligible contributions from
$|C|=2,3$ hadron-like states and they do not provide any independent constraint specific to 
our proposed model. The remaining four independent fourth order generalized charm susceptibilities,
$\chi_2^C,~\chi_{13}^{BC},~\chi_{22}^{BC}$ and $\chi_{31}^{BC}$ allow us the define the three
partial pressures, $p_q^C,~p_M^C$ and $p_B^C$ and one constraint
\begin{equation}
c_1 \equiv \chi_{13}^{BC} - 4\chi_{22}^{BC} + 3\chi_{31}^{BC} = 0,
\end{equation}
that has to hold if the model is correct.
If we consider the strange-charm sub-sector, we have six generalized susceptibilities
$\chi_{13}^{SC}$, $\chi_{22}^{SC}$, $\chi_{31}^{SC}$, $\chi_{112}^{BSC}$, $\chi_{121}^{BSC}$
and $\chi_{211}^{BSC}$. We can use three of these to estimate the partial pressures
$p_M^{C,S=1}$, $p_B^{C,S=1}$ and $p_B^{C,S=2}$ defined above, while the remaining ones
will provide three additional constraints that can used
to validate our proposed model. These constraints can be written as:
\begin{subequations}
\begin{eqnarray}
c_2 &\equiv& 2\chi_{121}^{BSC} + 4\chi_{112}^{BSC} + \chi_{22}^{SC} - 
2\chi_{13}^{SC} + \chi_{31}^{SC} = 0 , \\
c_3 &\equiv& 3\chi_{112}^{BSC} + 3\chi_{121}^{BSC} - \chi_{13}^{SC} + 
\chi_{31}^{SC}=0 , \\
c_4 &\equiv& \chi_{211}^{BSC} - \chi^{BSC}_{112}=0 .
\end{eqnarray}
\end{subequations}
Note that the above constraints hold trivially for a free charm-quark gas. It is assuring that 
our proposed model also smoothly connects to the HRG at $T_c$. In Fig.
\ref{fig:const} we show the lattice QCD data for $c_i$'s. Despite large errors on the
presently available lattice data all the $c_i$'s are, in fact, consistent with zero.
Note that, since a possible strange-charm di-quark-like excitation will carry
$|C|=|S|=1$ but $|B|=2/3$, the QCD data being consistent with the constraint $c_4=0$ actually 
tells us that the thermodynamic contributions of possible di-quark-like excitations are negligible in the 
deconfined phase of QCD.

One may speculate on the nature of these charm hadron-like excitations and, in
particular, why their partial pressures vanish gradually with increasing
temperature. A likely explanation may be that with increasing temperature the 
the spectral functions of these excitations gradually broaden. A detailed treatment
of thermodynamics of quasi-particles with finite width was developed in Refs.
\cite{Jakovac:2012tn,Biro:2014sfa,Jakovac:2013iua,Jakovac:2012tn}. It was shown that
broad asymmetric spectral functions lead to partial pressures that are considerably
smaller than those obtained with zero width quasi-particles of the same mass, and for
sufficiently large width the partial pressures can be made arbitrarily small. Thus,
the smallness of the partial pressure of charm quark-like excitations for $T\sim T_c$
may imply that they have a large width for those temperatures, while the widths of
the charm hadron-like excitations increase with the temperatures and these excitations become very
broad for $T\gtrsim200$ MeV. Such a gradual melting picture is also consistent with
the gradual changes of the screening correlators of open-charm meson-like excitations  with increasing 
temperature \cite{Bazavov:2014cta}.

Finally, one may wonder whether the rich structure of the up to fourth order
generalized charm susceptibilities can be described only in terms of the charm
quasi-quarks without invoking presence of any other type of charm degrees of freedom.
In terms of  charm quasi-quarks alone, the charm partial pressure will be 
$p^C/T^4=6/\pi^2 ~\hat m_c^2  K_2(\hat m_c)\cosh(\hat\mu_C+\hat\mu_B/3)$, where $\hat m_c=m_c/T$ with $m_c$
being the mass of the charm quasi-particle. The lattice QCD results for the charm
susceptibilities, for example, the non-vanishing values of $\chi_{mn}^{SC}$, can only
be described if the charm quasi-quark mass depends of the chemical potentials of all
the quark flavors, {\sl i.e.} $m_c\equiv m_c(T,\mu_B,\mu_S,\mu_C)$. For simplicity,
one may imagine Taylor expanding $m_c$ in terms of the chemical potentials and treat
these coefficients as parameters for fitting all the lattice QCD results on the
generalized charm susceptibilities. Obviously, such a quasi-quark model will contain
at least as many tunable parameters as the number of susceptibilities.  Moreover, in
order to satisfy various other constraints observed in the lattice QCD data, such
as  $\chi_4^C=\chi_2^C$, $\chi_{11}^{BC}=\chi_{13}^{BC}$, these parameters must also be very finely tuned. For
example, in order to satisfy the constraint $c_4=0$ the coefficients of the
$\mathcal{O}(\mu_B^2\mu_S\mu_C)$ term of $m_c$ must be equal to coefficient of the
$\mathcal{O}(\mu_B\mu_S\mu_C^2)$ term. Even if one chooses to use such finely tuned
parameters for the chemical potential dependence of the quasi-quark mass, the charm
partial pressure is not guaranteed to go smoothly over to the HRG values, as observed
in the lattice data.

To conclude, using the lattice QCD results for up to fourth order generalized charm
susceptibilities \cite{Bazavov:2014yba} we have shown that the weakly coupled charm
quasi-quarks becomes the dominant charm degrees of freedom only above $T\gtrsim~200$
MeV. To investigate the nature of charm degrees of freedom in the intermediate
temperature regime, $T_c\lesssim T\lesssim~200$ MeV, we postulated the presence of
non-interacting charm meson and baryon-like excitations in QGP, along with the charm
quark-like excitations.  We have shown that such a picture is consistent with the
presently available lattice QCD results. We have isolated the individual partial
pressures of these excitations and found that just above $T_c$ open-charm meson and
baryon-like excitations provide the dominant contribution to the thermodynamics of charm
sector. We also do not observe presence of di-quark like excitations in the $s$-$c$ sector 
at these temperatures. Our study hints at possible resonant scattering of the heavy quarks  
in the medium till around $1.2~T_c$ as first advocated in Ref. \cite{Mannarelli:2005pz}. 
These findings may have important consequences for the heavy quark
phenomenology of heavy-ion collision experiments, especially in understanding the
experimentally observed elliptic flow and nuclear modification factor of heavy
flavors at small and moderate values of transverse momenta 
\cite{Mannarelli:2005pz,Ravagli:2007xx,He:2012df,He:2014cla,Adil:2006ra,Sharma:2009hn}.

\emph{Acknowledgments:} This work was supported by U.S. Department of Energy under
Contract No. DE-SC0012704. The authors are indebted to the members of the
BNL-Bielefeld-CCNU collaboration for many useful discussions on this subject as well
as for sharing of the lattice QCD data. 

\bibliography{ref}

\end{document}